\documentclass{ifacconf}
\usepackage{amsmath}
\usepackage{graphics}
\usepackage{epsfig}
\usepackage{verbatim}
\usepackage{stmaryrd}
\usepackage{epstopdf}
\usepackage{xcolor}
\usepackage{color}
\usepackage{bbm}
\usepackage{amssymb}
\usepackage{mathrsfs}
\usepackage{caption}
\usepackage{nomencl}
\usepackage{multirow}
\usepackage{chngpage}
\usepackage{array}
\usepackage{comment}
\usepackage{enumerate}
\usepackage{graphicx}
\usepackage{transparent}
\usepackage{mathtools}
\usepackage{verbatim}
\usepackage{setspace}
\usepackage{courier}
\makenomenclature
\allowdisplaybreaks
\newcommand\rk{{\rm rank\,}} 
\newcommand\im{{\rm Im\,}}

\usepackage[normalem]{ulem} 

\usepackage{graphicx}      
\usepackage{natbib}        
\usepackage{tikz}
\usepackage{pgfplots}  
\usepackage{subfig}
\pgfplotsset{width=7cm, compat=newest}
\usepackage[siunitx,oldvoltagedirection]{circuitikz} 
\usetikzlibrary{calc}
\usetikzlibrary{shapes.geometric, arrows,patterns}
\usetikzlibrary{decorations.markings}
\usetikzlibrary{matrix}

\begin{document}
\begin{frontmatter}	
\title{An approximation for nonlinear differential-algebraic equations via singular perturbation theory}

\thanks[footnoteinfo]{This work was supported by Vidi-grant 639.032.733.}

\author[First]{Yahao Chen}
\author[First]{Stephan Trenn}

\address[First]{Bernoulli Institute for Mathematics, Computer Science, and
	Artificial Intelligence, University of Groningen, The Netherlands.}

\begin{abstract} 
In this paper, we study  jumps of nonlinear DAEs caused by inconsistent initial values. First, we   propose a simple normal form called the index-1 nonlinear Weierstrass form \textbf{(INWF)} for nonlinear DAEs. Then we generalize the notion of consistency projector introduced in \cite{LibeTren09}  for linear DAEs to the nonlinear case. {By an example, we compare our proposed nonlinear consistency projectors  with  two existing consistent initialization methods (one is from the paper \cite{LibeTren12} and the other is given by a MATLAB function) to show that   the two existing methods are not coordinate-free, i.e., the  consistent points calculated by the two methods are not invariant under nonlinear coordinates transformations.} Next we propose a singular perturbed system approximation for   nonlinear DAEs, which is an ordinary differential equation (ODE) with a small perturbation parameter, we show that the solutions of the proposed perturbation system approximate  both the jumps resulting from the nonlinear consistency projectors and the $\mathcal C^1$-solutions  of the DAE. At last, we use a numerical simulation  of a nonlinear DAE model arising from an electric circuit  to illustrate  the effectiveness of the proposed  singular perturbed system approximation of DAEs.

%
\end{abstract}
\end{frontmatter}
\section{Introduction}\label{sec:1}
\vspace{-0.3cm}
We consider a nonlinear differential-algebraic equation  (DAE),
\begin{align}\label{Eq:nonlinDAE}
\Xi:    E(x)\dot x=F(x),
\end{align}
where $x\in X$ is the vector of generalized states and $X$ is an open subset of $\mathbb R^n$, and where $E:X\to \mathbb R^{n\times n}$ and $F:X\to \mathbb R^{n}$ are $\mathcal C^{\infty}$-smooth maps. For each $x\in X$,  $E(x):T_xX\to \mathbb R^n$ is a linear map. A  DAE of the form (\ref{Eq:nonlinDAE}) will be denoted by $\Xi=(E,F)$ or $\Xi$. The matrix-valued function $E(x)$ is not necessarily invertible, which implies that there may exist  some algebraic constraints and some algebraic variables in the DAE $\Xi$. A particular case of $\Xi$ is a semi-explicit DAE
\begin{align}\label{Eq:seDAE}
\Xi^{SE}: \left\lbrace \begin{aligned}
\dot x_1&=  f_1(x_1,x_2),\\
0&=  f_2(x_1,x_2),
\end{aligned}\right.  
\end{align}
with $E=\left[ \begin{smallmatrix}
I_r&0\\
0&0
\end{smallmatrix}\right] $ being constant. The DAE $\Xi^{SE}$ has the algebraic variables  $x_2$  (since the derivatives of $x_2$ are not present) and the algebraic constraints $0=f_2(x_1,x_2)$. We will  study also linear DAEs of the form 
\begin{align}\label{Eq:linDAE}
\Delta: \  E\dot x=Hx,
\end{align}
where $E\in \mathbb R^{n\times n}$ and $H\in \mathbb R^{n\times n}$. A linear DAE of the form (\ref{Eq:linDAE}) will be denoted by $\Delta=(E,H)$ or, shortly, $\Delta$.  A linear DAE $\Delta$ is called \emph{regular} if   $sE-H\in \mathbb R^{n\times n}[s]\backslash0$.

	A $\mathcal C^1$-solution of a DAE $\Xi=(E,F)$ is a differentiable function $x:I\rightarrow X$ defined on an open interval $I$ such that for all $t\in I$, the curve $x(t)$ satisfies $E\left( {x(t)} \right)\dot x(t) = F\left( {x(t)} \right)$, where $\dot{x}$ denotes the classical time-derivative defined everywhere on $I$.
	A point $x^+_0$  is called  \emph{consistent}   if there exists at least one $\mathcal{C}^1$-solution  $x:I\to X$ with $t_0\in I$ such that  $x^+_0=x(t_0)$.
The set of all consistent points will be called consistency space and denoted by $S_c$. Without loss of generality, we can always assume $t_0=0$ and $I=(0,T)$ for some $T\in(0,\infty] $ (if not, we can re-parametrize the time variable $t$).  

  It is known that the $\mathcal C^1$-solutions of a nonlinear DAE $\Xi$ exist on its consistency space $S_c$ only (see Section \ref{sec:2}). For a given inconsistent initial point $x^-_0\in X\backslash S_c$, there does not exist any $\mathcal C^1$-solution starting from $x^-_0$.  Then it is natural to  search for the consistent point $x^+_0\in S_c$ such that we can get the $\mathcal C^1$-solutions of $\Xi$ starting from $x^+_0$. The instant change from the inconsistent point  $x^-_0$
  to a consistent one $x^+_0$
  is called a jump of the DAE  at $t = 0$.   Note that the jumps  which  we study in the paper   are called  external or exogenous  jumps, which are different from the jumps  at the impasse (or singular) points as discussed  in  \cite{takens1976constrained,chua1989impasse,SastDeso81}. We assume throughout that once starting from the point $x^+_0$, there will not exist any jump and we will study only the $\mathcal C^1$-solutions of $\Xi$.  In conclusion, we consider the following initial value problem:
$$
\left\lbrace \begin{array}{l} 
{\rm Jumps:\ } \lim\limits_{t\to 0^-} x(t)=x^-_0\notin S_a\to \lim\limits_{t\to 0^+} x(t)=x^+_0\in S_a,\\
\mathcal C^1\text{-solutions: }\left( E(x)\dot x\right) _{(0,T)}=F(x)_{(0,T)},
\end{array}\right. 
$$
for some function $x:I\to\mathbb{R}^n$ differentiable on $(0,T)\subset I$. The problem of finding the consistent point $x^+_0$ for a DAE with an inconsistent initial value $x^-_0$ is called consistent initialization, which is a significant problem for hybrid DAE systems involving with jump behaviors. Some examples of such systems are  the electric circuits with instant connections or switching devices  (see e.g., \cite{zuhao1991zz,vlach1995analysis,Tren12}), the power systems with DC transmissions in \cite{susuki2008discontinuous}, the multi-body dynamics in \cite{hamann2008numerical}  and  the battery model of \cite{METHEKAR20112227}. 

 For a regular linear DAE $\Delta=(E,H)$, given by (\ref{Eq:linDAE}),  the consistent initialization can be solved by the linear consistency projector 
 introduced by \cite{LibeTren09,LibeTren12}, which is a linear map constructed with the help of the well-known Weierstrass form \textbf{(WF)}. 
 For a semi-explicit DAE $\Xi^{SE}$ of the form (\ref{Eq:seDAE}),  the singular perturbation theory  (see e.g., \cite{KokoKhal99,Khal01}) was frequently  used    to  study   system approximations of  the discontinues solutions of $\Xi^{SE}$  (see e.g., \cite{SastDeso81,RabiRhei02,susuki2008discontinuous} and Section \ref{sec:4} of the present paper).
   {Two existing methods of solving the consistent initialization problem for nonlinear DAEs are, the  jump rule of \cite{LibeTren12}, which determines the consistent initial value $x^+_0$ through the formula $x^+_0-x^-_0\in\ker E(x^+_0)$, and the function \emph{decic} of MATLAB (see \cite{matdecic}), which  calculates the consistent initial values via a numerical searching method, we will show in Example \ref{Ex:decicly} below that both of those two methods are not coordinate-free, i.e.,  the calculated consistent values depends on which local coordinates are chosen for the DAE.}
 
The aims of this paper are, on one hand, to give a nonlinear generalization of the linear consistency projector in order to calculate consistent initial points for nonlinear DAEs, on the other hand, to extend the singular perturbed system approximation method to nonlinear DAEs of the form (\ref{Eq:nonlinDAE}) to study the jump behaviors.   This paper is organized as follows: We introduce the notations of the paper   and some  notions as invariant submanifolds, external equivalence and linear consistency projectors  in Section \ref{sec:2}. We propose a normal form called the index-1  nonlinear Weierstrass form \textbf{(INWF)} and extend the linear consistency projector to nonlinear DAEs in Section \ref{sec:3}. A singular perturbed system approximation of nonlinear DAEs is proposed in Section \ref{sec:4} and we show  the simulation result of our  singular perturbation method  applied to an electric circuit in Section \ref{sec:5}. Conclusions  are given in Section \ref{sec:6}.  
\section{Notations and some preliminaries of nonlinear DAEs}\label{sec:2}
We  use the following notations: The symbol  $\mathcal C^k$ denotes the class of functions which are $k$-times differentiable.  For a map $A:X\to \mathbb R^{n\times n}$, $\ker A(x)$, ${\rm Im\,} A(x)$ and ${\rm rank\,}A(x)$ are the kernel, the image and the rank of $A$ at $x$, respectively. The general linear group over $\mathbb R$ of degree $n$ is denoted by $GL(n,\mathbb R)$.  For two column vectors  $v_1\in \mathbb R^m$ and $ v_2\in \mathbb R^n$, we  write $(v_1,v_2)=[v^T_1,v^T_2]^T\in \mathbb R^{m+n}$.  Let $f_i:X\to \mathbb R$ for $i=1,\dots,m$, in coordinates $x=(x_1,\dots, x_n)$, the differential of $f_i$ is ${\rm d} f_i=\sum^n_{j=1}\frac{\partial f_i}{\partial x_j}{\rm d}x_j=[\frac{\partial f_i}{\partial x_1},\dots,\frac{\partial f_i}{\partial x_n}]$, the differentials of a vector-valued function $f=(f_1,\dots,f_m)$ are ${\rm D} f =\left[ \begin{smallmatrix}
{\rm d} f_1\\
\vdots\\
{\rm d} f_m
\end{smallmatrix}\right]  $. 
We assume that the reader is familiar with some basic notions as smooth  embedded submanifolds, tangent spaces, involutive distributions from differential geometry, the reader can also consult the book by \cite{lee2001introduction} for the definitions of such notions.

The existence and uniqueness of $\mathcal C^1$-solutions for nonlinear DAEs of the form (\ref{Eq:nonlinDAE}) have been discussed using geometric methods   in e.g., \cite{reich1991,RabiRhei02,chenMTNS,chennonlinear2020}. An important notion in the geometric solutions theory of DAEs is the invariant submanifold  defined as follows.
\begin{defn}
	For a DAE $\Xi=(E,F)$,  a smooth connected embedded submanifold  $M$  is called  \emph{invariant} if   for any $ x^+_0\in M$, there exists a $\mathcal C^1$-solution $x: I\to X$ such that $x(t_0)=x^+_0$ with $t_0\in I$ and $ x(t)\in M$, $\forall\, t\in I$.
	Fix a  point $x_p\in X$, a smooth embedded submanifold $M$ containing $x_p$ is called \emph{locally invariant},   if there exists a neighborhood $U$ of $x_p$ such that $M\cap U$ is invariant.
\end{defn}
A locally invariant submanifold $M^*$, around a point $x_p$, is called locally \emph{maximal}, if there exists a neighborhood $U$ of $x_p$ such that for any other locally invariant submanifold $M$, we have $M\cap U \subseteq M^*\cap U$. It is  shown in \cite{chenMTNS,chennonlinear2020} that the maximal invariant submanifold $M^*$ around a nominal point $x_p$ locally coincides with the consistency space $S_c$, i.e., there exists a neighborhood  $U^*$ of $x_p$ such that $$M^*\cap U^*=S_c\cap U^*.$$ Hence in the present paper, we make no difference between the notion of maximal invariant submanifold $M^*$ and that of   consistency space $S_c$ when considering a DAE $\Xi$ around a point $x_p$. Note that there is an iterative way of calculating the locally maximal invariant submanifold $M^*$ of DAEs, called the geometric reduction method (see e.g., \cite{RabiRhei02,chenMTNS,chennonlinear2020}), the number of steps for the geometric reduction method to produce $M^*$ and to get the solutions of a DAE is called the \emph{geometric} index (see \cite{chenMTNS}) of the DAE.

We now recall a definition of equivalence for linear DAEs,
 two linear DAEs $\Delta=(E,H)$ and $\tilde \Delta=(\tilde E,\tilde H)$ are called externally equivalent (see \cite{chen2020}) or  strictly equivalent if there exist constant and invertible matrices $Q$ and $P$ such that $\tilde E=QEP^{-1}$ and $\tilde H=QHP^{-1}$.  The same concept can be generalized to nonlinear DAEs of form  (\ref{Eq:nonlinDAE}) as follows.
\begin{defn}[external equivalence]\label{Def:ex-equi}
Consider two DAEs $\Xi=(E,F)$ and $\tilde \Xi=(\tilde E,\tilde F)$ defined on $X$ and $\tilde X$, respectively. Then  $\Xi$ and $\tilde \Xi$ are called externally equivalent, shortly ex-equivalent, if there exist a diffeomorphism $\psi: X\rightarrow \tilde X$ and    $Q: X\rightarrow GL(n,\mathbb{R})$ such that
	$$
	\tilde E(\psi (x)) \!=\!Q(x)E(x)\left( \frac{\partial \psi (x)}{\partial x}\right) ^{-1}, \ \
	\tilde F(\psi (x))\! =\!Q(x)F(x).
	$$
	The ex-equivalence of two DAEs will be denoted by $\Xi\mathop  \sim \limits^{ex}\tilde \Xi$. If  $\psi: U\rightarrow \tilde U$ is a local diffeomorphism between neighborhoods $U$ of $x_p$ and $\tilde U$ of $\tilde x_p$, and $Q(x)$ is defined on $U$,  we will speak about local ex-equivalence.
\end{defn}
\begin{rem}
It is easily seen, that for two externally equivalent systems $\Xi$ and $\tilde{\Xi}$ a $\mathcal C^1$-curve $x:I\to X$ is a solution of $\Xi$ if and only if $\psi\circ x$ is a solution of $\tilde \Xi$. 
\end{rem} 
To illustrate the notions of maximal invariant submanifold and external equivalence, we use the following example.
\begin{exmp}\label{Ex:decicly0} 
	Consider a DAE $\Xi=(E,F)$, given by
	\begin{align}\label{Eq:exmIFJ}
	\Xi: \left[ \begin{matrix}
	1&3x^2_2-1\\
	0&0
	\end{matrix} \right]  \left[ \begin{matrix}
	\dot x_1\\
	\dot x_2
	\end{matrix} \right]=\left[ \begin{matrix}
	- x_2\\
	x_1
	\end{matrix} \right].
	\end{align} 
	Fix a point $x_p=(x_{1p},x_{2p})=(0,1)$, the locally maximal invariant submanifold of $\Xi$ around $x_p$ is    
	$M^*=\left\lbrace x\in \mathbb R^2\,|\, x_1=0, x_2>\frac{\sqrt{3}}{3}\right\rbrace $ (note that $M^*$ is connected).  We have that $\Xi$ is locally ex-equivalent to the following form (i.e., the $\mathbf{(INWF)}$, see Definition \ref{Def:inwf}) 
	\begin{align}\label{Eq:inwfex}
	\left[\begin{matrix}
	1&0\\
	0&0
	\end{matrix} \right] \left[\begin{matrix}
	\dot {\xi}_1\\
	\dot {\xi}_2
	\end{matrix} \right]= \left[\begin{matrix}
	-f(\xi_1,0)\\
	\xi_2
	\end{matrix} \right],
	\end{align}
	on the neighborhood $V=\left\lbrace x\in \mathbb R^2\,|\, x_2>\frac{\sqrt{3}}{3}\right\rbrace $ of $x_p$,  via $\psi=\xi=(\xi_1,\xi_2)=(x_1+x_2^3-x_2, x_1)$	 and $Q=\left[ \begin{smallmatrix}
	1&-f'\\
	0&1
	\end{smallmatrix}\right] $, where $f(\xi)=f(\xi_1,0)+f'(\xi)\xi_2$, $f=	\frac{1}{3}\left( a  \! +\! (a^2\! -\! \frac{1}{27})^{\frac{1}{2}}\right) ^{-\frac{1}{3}}\! +\! \left( a  \! +\! (a^2 \!-\! \frac{1}{27})^{\frac{1}{2}}\right) ^{\frac{1}{3}}$,
	$a(\xi_1,\xi_2)=\frac{\xi_1-\xi_2}{2}
	$.  
\end{exmp} 
\section{Index-1 nonlinear Weierstrass form and nonlinear consistency projector}\label{sec:3}
Consider a nonlinear DAE $\Xi=(E,F)$, let $H(x,\dot x)=E(x)\dot x -F(x)$, define the $k$-th order differential array of $H(x,\dot x)=0$ by
\begin{align}\label{Eq:darray} 
H_k(x,x',w)\!=\!\left[ \begin{smallmatrix} 
H\\
{\rm D}_xHx'+{\rm D}_{x'}Hx''\\
\vdots\\
\frac{{\rm d}^k}{{\rm d}t^{k}}H
\end{smallmatrix}  \right] (x,x',w)\!=\!0,
\end{align}
where $w= \left( x^{(2)},\dots,x^{(k+1)}\right) $, the differentiation index   or shortly, the index,  of the DAE $\Xi$ is the least integer $k$  such that equation (\ref{Eq:darray}) uniquely determines $x'$ as a function of $x$, i.e., $x'=v(x)$. In  \cite{chenMTNS}, we have shown that under some constant rank assumptions, the differential index coincides with the geometric index, we will  use a simplification of those constant rank assumptions in the present paper: For a DAE $\Xi=(E,F)$, fix a point $x_p$, define $F_2:=F\backslash \im E=Q_2F$, assume that $F_2(x_p)=0$  and  introduce the following constant rank condition, there exists a neighborhood $U$ of $x_p$ such that
\begin{description}
	\item[\textbf{(CR)}]  $\rk E(x)=const.$, $\forall x\in U$;   $\rk {\rm D}F_2(x)=const.$ and $\rk (E\ker {\rm D}F_2(x))=const.$, $\forall x\in U $ such that $ F_2(x)=0$. 
\end{description} 
	The assumption $\rk E(x)=const.$ ensures that there exists $Q:U\to GL(n,\mathbb R)$ such that $E_1$ of $QE=\left[ \begin{smallmatrix}
E_1\\
0
\end{smallmatrix}\right] $ is of full row rank. Denote $QF=\left[ \begin{smallmatrix}
F_1\\
F_2
\end{smallmatrix}\right]$, then the map $F\backslash\im E$ is given by $F_2$. The assumption $\rk {\rm D}F_2(x)=const.$ guarantees that the zero-level set $\left\lbrace x\in U\,|\, F_2(x)=0\right\rbrace $ is a smooth embedded submanifold and the condition $\rk (E\ker {\rm D}F_2(x))=const.$ excludes singular/impasses points (see  \cite{chua1989impasse,chen2019geometric}) and helps to view the DAE as an ODE defined on a submanifold. Note that under the condition \textbf{(CR)}, a DAE $\Xi$ is of differentiation index-1 if and only if it is of geometric index-1 (\cite{chenMTNS}). 
Now we define a normal form, which is a semi-explicit DAE of index-1 with the  algebraic equations fully decoupled from its differential equations.
\begin{defn}[index-1 nonlinear Weirstrass form]\label{Def:inwf}
	We say that a DAE $\Xi$ is represented in the index-1 nonlinear Weirstrass form \textbf{(INWF)}  if $\Xi$ is of the form
	\begin{align}\label{Eq:inwf}
\left\lbrace 	\begin{aligned}
\dot \xi_1&=F^*(\xi_1),\\
0&=\xi_2.
	\end{aligned}\right. 
	\end{align}
	where $\xi_1\in X_1\subseteq \mathbb R^r$, $\xi_2\in X_2\subseteq \mathbb R^{n-r}$ and $F^*: X_1\to \mathbb R^r$.
\end{defn}
\begin{rem}\label{rem:niwf}
For any DAE in \textbf{(INWF)} with an inconsistent initial point $(\xi^-_{10},\xi^-_{20})\notin M^*$, i.e., $\xi^-_{20}\neq 0$ (it is clear that the maximal invariant submanifold of (\ref{Eq:inwf}) is $M^*=\left\lbrace (\xi_1,\xi_2)\in X_1\times X_2 \,|\, \xi_2=0\right\rbrace $), we could easily deduce that $(\xi^+_{10},\xi^+_{20})=(\xi^-_{10},0)$ is the only possible jumping point from $(\xi^-_{10},\xi^-_{20})$. Indeed, for the DAE (\ref{Eq:inwf}),  only $\xi_2$-variables are allowed to jump  because  any jump of $\xi_1$-variables will produce a Dirac impulse  on the left-hand side of $\dot \xi_1=F^*(\xi_1)$ (see the distributional solution theory of DAEs in \cite{Tren09b}), which is not possible since $F^*(\xi_1)$ is not able to produce a same impulsive term on the right-hand side in order to  equalize  the differential equations.
\end{rem}
\begin{thm}
Consider a DAE $\Xi=(E,F)$ and fix a point $x_p\in X$. Assume that $\Xi$ satisfies the  condition \textbf{(CR)} in a  neighborhood $U\subseteq X$ of $x_p$. Then there exists a neighborhood $V\subseteq U$ of $x_p$ such that $\Xi$ is locally ex-equivalent to the \textbf{(INWF)}, given by (\ref{Eq:inwf}), if and only if $\Xi$ is of index-1 and the distribution $\mathcal E=\ker E$ is involutive. 
\end{thm}
\begin{pf} 
	\emph{Only if.} Assume that $\Xi$ is locally ex-equivalent to the   \textbf{(INWF)}, denoted by $\tilde \Xi=(\tilde E,\tilde F)$. It is clear that $\tilde \Xi$ is index-1 and that $\ker \tilde E$ is involutive (since  $\tilde E$  is constant).  
	Notice that the $Q$-transformation preserves the kernels and $\ker \tilde E(\psi(x))=\frac{\partial \psi}{\partial x}\ker E(x)$; let $\ker E={\rm span}\left\lbrace g_1,\dots,g_{n-r} \right\rbrace $ for some vector fields $g_i$, we have $\ker \tilde E={\rm span}\left\lbrace \frac{\partial \psi}{\partial x}g_1,\dots,\frac{\partial \psi}{\partial x}g_m \right\rbrace$, so the Lie bracket $[g_i,g_j]\in \ker E$ (i.e., $\ker E$ is involutive) if and only if  $[\frac{\partial \psi}{\partial x}g_i,\frac{\partial \psi}{\partial x}g_j]=\frac{\partial \psi}{\partial x}[g_i,g_j]=\frac{\partial \psi}{\partial x}\ker E=\ker \tilde E$ (i.e., $\ker \tilde E$ is involutive). We conclude that $\Xi$  is index-1 and $\mathcal E=\ker E$   is involutive as well. 

	\emph{If.}	  Suppose that $\Xi$ is of index-1 and the distribution $\mathcal E=\ker E$ is involutive. Then by  $\rk E(x)=const.$ (denote this rank by $r$) of \textbf{(CR)},  there exists $Q:U\to GL(n,\mathbb R)$ such that $\rk E_1(x)=r$ in 
	\begin{align}\label{Eq:QDAE}
	Q(x)E(x)\dot x=Q(x)F(x) \Rightarrow \left[ \begin{smallmatrix}
E_1(x)\\
0
\end{smallmatrix}\right]\dot x =\left[ \begin{smallmatrix}
F_1(x)\\
F_2(x)
\end{smallmatrix}\right].
	\end{align}
 Notice that the condition \textbf{(CR)}	implies that there exists a neighborhood $U_1\subseteq U$ of $x_p$ such that $\rk A(x)=\rk \left[ \begin{smallmatrix}
E_1(x)\\
{\rm D}F_2(x)
\end{smallmatrix}\right]=const.$, $\forall x\in U_1: F_2(x)=0$. Since the DAE is of differentiation  index-1, we have that $A(x)$ has to be invertible, i.e.,  $\rk A(x)=n$, because only if  $A(x)$ is invertible, we can uniquely solve $\dot x=v(x)=A^{-1}(x) \left[ \begin{smallmatrix}
E_1(x)\\
{\rm D}F_2(x)
\end{smallmatrix}\right]$ with only a first order differentiation of (\ref{Eq:QDAE}) (note that we only need to differentiate the algebraic equation $0=F_2(x)$). Since the distribution $\Xi=\ker E$ is involutive, by Frobenius theorem (see e.g., \cite{lee2001introduction}),   there exist a neighborhood $U_2\subseteq U_1$ and a smooth map $\xi_1:U_2\to \mathbb R^{r}$ such that ${\rm span}\left\lbrace {\rm d}  \xi^1_1, \dots,  {\rm d}  \xi^r_1\right\rbrace =\mathcal E^{\bot}$, where ${\rm d} \xi^i_1$ are independent rows of ${\rm D}\xi_1$ and $\mathcal E=\ker E=\ker E_1$, i.e., ${\rm D}\xi_1(x)\ker E_1(x)=0$, $\forall x\in U_2$. It follows that there exists $Q_1:U_2\to GL(r,\mathbb R)$ such that ${\rm D}\xi_1(x)=Q_1(x)E_1(x)$.   Set $\xi_2=F_2$, then we have $\psi(x)=(\xi_1(x),\xi_2(x))$ is a local diffeomorphism on $U_2$ since 
$$
\frac{\partial \psi(x)}{\partial x}=\left[ \begin{smallmatrix}
{\rm D}\xi_1(x)\\
{\rm D}F_2(x)
\end{smallmatrix}\right] =\left[ \begin{smallmatrix}
Q_1(x)&0\\
0&I
\end{smallmatrix}\right]\left[ \begin{smallmatrix}
E_1(x)\\
{\rm D}F_2(x)
\end{smallmatrix}\right]=\left[ \begin{smallmatrix}
Q_1(x)&0\\
0&I
\end{smallmatrix}\right]A(x)
$$ is invertible for all $x\in U_2$.  Define the new local coordinates $\xi=\psi=(\xi_1,\xi_2)$ on $U_2$, the DAE (\ref{Eq:QDAE}) under the new $\xi$-coordinates is represented by 
$$
	\left[ \begin{smallmatrix}
E_1(x)\\0
\end{smallmatrix}\right] \left( \frac{\partial \psi(x)}{\partial x}\right)^{-1}  \frac{\partial \psi(x)}{\partial x}\dot x=\left[ \begin{smallmatrix}
F_1(x)\\F_2(x)
\end{smallmatrix}\right]\Leftrightarrow 
$$  
	\begin{align}\label{Eq:DAE4}
	\left[\begin{smallmatrix}
E^1_1(\xi_1, \xi_2)&0\\
0&0
\end{smallmatrix} \right] \left[\begin{smallmatrix}
\dot {\xi}_1\\
\dot {\xi}_2
\end{smallmatrix} \right]= \left[\begin{smallmatrix}
\tilde F_1({\xi}_1,{\xi}_2)\\
\xi_2
\end{smallmatrix} \right], 
	\end{align}
	where $E^1_1:U_2\to \mathbb R^{r\times r}$, $[E^1_1\circ \psi, E^2_1\circ \psi]= E_1(\frac{\partial \psi}{\partial x})^{-1}$ with $E^2_1\equiv0$, $\tilde F_1\circ \psi=F_1$. Notice that $E^2_1=0$ because $\im E^2_1(x)=E_1(x)\ker {\rm D}\xi_1(x)=0$ and that  $E^1_1(x)$ is invertible for $x\in U_2$ since $\rk E(x)=const.=r$, $\forall x\in U_2$. Let $\bar F_1 = (E^1_1)^{-1}\tilde F_1$,  we can always find $\bar F'_1:U_2\to \mathbb R^{r\times m}$ such that $\bar F_1({\xi}_1,{\xi}_2)=\bar F_1({\xi}_1,0)+\bar F'_1(\xi_1,\xi_2)\xi_2$. Then via $\tilde Q=\left[ \begin{smallmatrix}
	(E^1_1)^{-1}&-\bar F'_1\\
	0&I
	\end{smallmatrix}\right] $, the DAE (\ref{Eq:DAE4}) is  ex-equivalent to the \textbf{(INWF)} with $F^*(\xi_1)=\bar F_1(\xi_1,0)$. Finally, it is seen that $\Xi$ is locally (on $V=U_2$) ex-equivalent to the \textbf{(INWF)} via the diffeomorphism $\psi$ and  the $\tilde QQ$-transformation.
\end{pf}
With the help of the \textbf{(INWF)}, we can generalize the notion of consistency projector to nonlinear DAEs:
\begin{defn}[nonlinear consistency projector]
	For a nonlinear DAE $\Xi=(E,F)$, fix a point $x_p$ and assume that there exists a neighborhood $V$ of  $x_p$ such that $\Xi$ is locally (on $V$) ex-equivalent to the \textbf{(INWF)}, given by  (\ref{Eq:inwf}), via a $Q$-transformation and a local diffeomorphism $\psi$. The  (local) \emph{nonlinear consistency projector} $\Omega_{E,F}:V\backslash M^*\to V\cap M^*$ of $\Xi$ is then defined by 
	$$
	\Omega_{E,F}:=\psi^{-1}\circ\pi\circ\psi, 
	$$
	where $\pi:\mathbb R^n\to \mathbb R^n$ is the canonical projection  attaching {$(\xi_1,\xi_2)\mapsto (\xi_1,0)$}.
\end{defn}
For a DAE $\Xi$ being locally (on $V$) ex-equivalent to the \textbf{(INWF)}  with an inconsistent initial value $x^-_0\in  V\backslash M^*$, we can get a unique consistent point $x^+_0=\Omega_{E,F}(x^-_0)\in V\cap  M^*$ since in the $\xi$-coordinates of the \textbf{(INWF)}, the inconsistent point $(\xi^-_{10},\xi^-_{20})=\psi(x^-_0)$ has to jump into $(\xi^+_{10},\xi^+_{20})=(\xi^-_{10},0)$ (see Remark \ref{rem:niwf}), hence $x^+_0=\psi^{-1}(\xi^+_{10},\xi^+_{20})=\psi^{-1}\circ\pi\circ\psi(x^-_0)=\Omega_{E,F}(x^-_0)$. {Then we compare the consistent initial values calculated by the nonlinear consistency projector with that from the jump rules in \cite{LibeTren12} and   MATLAB \emph{decic} function (see \cite{matdecic}).}
\begin{exmp} [continuation of \emph{Example \ref{Ex:decicly0}}]\label{Ex:decicly} 
  The DAE (\ref{Eq:exmIFJ})  satisfies the condition \textbf{(CR)} in the neighborhood $U=\left\lbrace x\in \mathbb R^2\,|\, x\neq \pm \sqrt{3}/{3}\right\rbrace $ of $x_p$.  We have shown in
 Example~\ref{Ex:decicly0} that (\ref{Eq:exmIFJ}) is ex-equivalent (on $V\subseteq U$) to the \textbf{(INWF)}, given by (\ref{Eq:inwfex}), via $Q$ and $\psi$. Thus the  nonlinear (local) consistency projector of $\Xi$ is
	$$
	\Omega_{E,F}=\psi^{-1}\circ\pi\circ\psi= \left[ \begin{matrix}
	0\\
	f(x_1+x^2_2-x_2,0)
	\end{matrix}\right].
	$$
Take an inconsistent initial value $x^-_0=(1,0.7)\in V\backslash M^*$, the consistent point calculated by the nonlinear consistency projector is  $x^+_0=\Omega(x^-_0)=(0,1.233)\in M^*$. Note that the inconsistent initial point of (\ref{Eq:inwfex}) is  $\xi^-_0=\psi(x^-_0)=(0.643,1)$ and the consistent point  is $\xi^+_0=(0.643,0)$ since only $\xi_2$-variables are allowed to jump (see Remark \ref{rem:niwf}).
 Then we use the jump rule $x^+_0-x^-_0=\ker E(x^+_0)$ in \cite{LibeTren12} to calculate the consistent values $\tilde x^+_0$  and $\tilde \xi^+_0$ for (\ref{Eq:exmIFJ}) and (\ref{Eq:inwfex}), respectively, and we get 
	$$
	\tilde x^+_0=(0, 0.109)  \text{ and } \tilde \xi^+_0=(0.643,0).
	$$
	Similarly, we use MATLAB \emph{decic} function to determine  the consistent values $\bar x^+_0$  and $\bar \xi^+_0$ for  (\ref{Eq:exmIFJ}) and (\ref{Eq:inwfex}), respectively, to get 
	$$
	\bar x^+_0=(0, 0.7)  \text{ and } \tilde \xi^+_0=(0.643,0).
	$$
	Since $\tilde \xi^+_0\neq \psi(\tilde x^+_0)$ and $\bar \xi^+_0\neq \psi(\bar x^+_0)$, we conclude that the two consistent initialization methods in \cite{LibeTren12} and \cite{matdecic} do \emph{not} preserve the calculated consistent points when changing the coordinates of the given DAE.	On the other hand, the jump $x^-_0\to x^+_0$ of (\ref{Eq:exmIFJ}), given by the nonlinear consistency projector, and the jump
	 $\xi^-_0\to \xi^+_0$ of (\ref{Eq:inwfex}) are clearly the same jump in different coordinates since $\xi^-_0=\psi{(x^-_0)}$, $\xi^+_0=\psi{(x^+_0)}$, which proves that  the consistent initialization calculated by the consistency projector is coordinate-free.
\end{exmp}
 \section{Singular perturbed system approximation of nonlinear DAEs}\label{sec:4}
We first recall a singular perturbed system for a semi-explicit DAE $\Xi^{SE}$ of the form (\ref{Eq:seDAE}).   Replacing the algebraic constraint $0=f_2(x_1,x_2)$ by $\epsilon \dot x_2=f_2(x_1,x_2)$, where $\epsilon$   represents some  modeling parameters which can be ignored (e.g,  the small inductance of an inductor in electrical circuits, see page 367 of \cite{RabiRhei02}),  we get a  perturbed ODE system $\Xi^{SE}_{\epsilon}$ on the left-hand side of the following formula, then by rescaling time $t$ to $\tau$ by $\frac{d\tau}{dt}=\frac{1}{\epsilon}$, we get a perturbed system in the   time-scale $\tau$ on the right-hand side.
 \begin{align*}
 \Xi_{\epsilon}^{SE}: \left\lbrace \begin{aligned}
 \dot x_1&=  f_1(x_1,x_2),\\
 \epsilon\dot x_2&=  f_2(x_1,x_2).
 \end{aligned}\right.  \overset{\epsilon=\frac{dt}{d\tau}}{\Leftrightarrow}	\left\lbrace \begin{aligned}
 \frac{d x_1}{d  \tau}&=\epsilon f_1(x_1,x_2),\\
 \frac{d x_2}{d  \tau}&=f_2(x_1,x_2).
 \end{aligned}\right. 
 \end{align*} 
There are, in general, two assumptions in the   singular perturbed approximation method of semi-explicit DAEs: (a) $\frac{d f_2}{d x_2}$ is invertible (which is actually equivalent to that $\Xi^{SE}$ is of index-1); (b) the so-called boundary layer model $\frac{d x_2}{d \tau}=f_2(x^-_{10},x_2)$ is asymptotically stable  uniformly in $x_2$. Then under assumptions (a),(b), the well-known Tihkonov's theorem (see e.g., \cite{Khal01} and  a similar result  in Theorem III.1 of \cite{SastDeso81}) states that if a unique solution $(x_1(t),x_2(t))$ of $\Xi^{SE}$ starting from a consistent initial point $(x^{+}_{10},x^+_{20})$ exists on the interval $I=(0,\alpha)$, then there exists $\delta\ge 0$ such that a solution $(\bar x_1(t,\epsilon),\bar x_2(t,\epsilon)) $ of $\Xi_{\epsilon}^{SE}$ starting from any point $(x^-_{10},x^-_{20})$ with $||x^{+}_{10}-x^{-}_{10}||+||x^{+}_{20}-x^{-}_{20}||<\delta$ satisfies
 \begin{align}\label{Eq:conv}
 \begin{aligned}
 \lim\limits_{\epsilon\to 0} || x_1(t)-\bar x_1(t,\epsilon)||&=0, \\ \lim\limits_{\epsilon\to 0} || x_2(t)-\bar x_2(t,\epsilon)||&=0,
 \end{aligned}
\end{align}
 on all closed subintervals of $I$.  In this section, we will propose a singular perturbed system approximation for nonlinear DAEs of the form (\ref{Eq:nonlinDAE})    with the help of the proposed normal form \textbf{(INWF)}.
 \begin{defn}[singular perturbed system]\label{Def:sps}
 For a nonlinear DAE $\Xi=(E,F)$, fix a point $x_p$, assume that there exists a neighborhood $V$ of  $x_p$ such that $\Xi$ is locally (on $V$) ex-equivalent to   the \textbf{(INWF)} of (\ref{Eq:inwf})   via a $Q$-transformation and a local diffeomorphism $\psi$.  
 	Define the following singular perturbed system on $V$:
 	\begin{align}\label{Eq:ps}
 	\Xi_{\epsilon}: \dot x = E_{\epsilon}^{-1}(x, \epsilon)F(x) , 
 	\end{align} 
 	where $E_{\epsilon} (x, \epsilon)=E(x)+Q^{-1}(x)\left[ \begin{matrix}
 	0&0\\
 	0&-\epsilon I_{n-r}
 	\end{matrix}\right] \frac{\partial \psi(x)}{\partial x}$.
 \end{defn}
 \begin{rem}
 	Any linear index-1 regular DAE $\Delta=(E,H)$ of the form (\ref{Eq:linDAE})  is always ex-equivalent to a decoupled DAE given by $\left(\left[ \begin{smallmatrix}
 	I_{n_1}&0\\ 0&0	\end{smallmatrix}\right],\left[ \begin{smallmatrix} A_1&0\\ 0&I_{n_2} \end{smallmatrix}\right]\right)$. Applying the construction of (\ref{Eq:ps}) to $\Delta$, we get the following singular perturbed system:
 	$$
 	\Delta_{\epsilon}: \dot x = E_{\epsilon}^{-1}Hx=P^{-1} \left[ \begin{smallmatrix}
 	A_1&0\\
 	0&-\frac{1}{\epsilon} I_{n_2}
 	\end{smallmatrix}\right] Px,
 	$$
 	where $ E_{\epsilon} =Q^{-1}\left[ \begin{smallmatrix}
 	I_{n_1}&0\\
 	0&-\epsilon I_{n_2}
 	\end{smallmatrix}\right] P$.
The above perturbed linear system $\Delta_{\epsilon}$ is proposed in Section IV of \cite{mironchenko2015} as an ODE approximation of linear DAEs.
 \end{rem}
 The following theorem shows  that the solution $\bar x(t,\epsilon)$ of  the proposed   perturbed system $\Xi_{\epsilon}$ of (\ref{Eq:ps}) with an inconsistent initial value $x^-_0$ converges to the $\mathcal C^1$-solution $x(t)$ of $\Xi$  staring from a consistent point $x^+_0$ calculated via the nonlinear consistency projector.  
 \begin{thm}\label{Thm:sps}
 	Consider a  DAE $\Xi=(E,F)$  and fix a  point $x_p\in X$.   Assume that   the  condition   \textbf{(CR)} is satisfied in a neighborhood $U$ of $x_p$. Suppose that $\Xi$ is of geometric index-1 and that $\mathcal E=\ker E$ is involutive, implying  that   there exists a neighborhood $V\subseteq U$ of $x_p$ such that $\Xi$ is locally (on $V$) ex-equivalent to  the \textbf{(INWF)} of (\ref{Eq:inwf}) via $Q$ and $\psi$. Let $x^-_0\in V\backslash M^*$ be an inconsistent initial point of $\Xi$ and $x^+_0=\Omega_{E,F}(x^-_0)\in M^*$ be the consistent point calculated via the nonlinear consistency projector $\Omega_{E,F}$. If $\bar x(t,\epsilon): I\to V$ is the solution of the perturbed system $\Xi_{\epsilon}$ of (\ref{Eq:ps})  starting from  $x^-_0$  and $x(t):I\to V$ is the $\mathcal C^1$-solution of $\Xi$ starting from $x^+_0$, then we have 
 	\begin{align}\label{Eq:Jconvg}
 	\lim\limits_{\epsilon\to 0} ||\bar x(t,\epsilon)-x(t)||=0, \ \ \forall t\in I.
 	\end{align}  
 \end{thm}
\begin{pf}
	Suppose that $\Xi$ is locally (on $V$) ex-equivalent to the \textbf{(INWF)} of (\ref{Eq:inwf}) via $Q$ and $\psi$. Consider the following disturbed system for (\ref{Eq:inwf}):
\begin{align}\label{Eq:rsp}		 \left[\begin{matrix}
\dot \xi_1\\
\dot \xi_2
\end{matrix} \right]= \left[\begin{matrix}
I_{r}&0\\
0& -\epsilon I_{n-r}
\end{matrix} \right]^{-1}\left[\begin{matrix}
F^*(\xi_1)\\
\xi_2
\end{matrix} \right]=\left[\begin{matrix}
F^*(\xi_1)\\
-\frac{1}{\epsilon}\xi_2
\end{matrix} \right], 
\end{align} 
Let $\bar \xi(t,\epsilon)=(\bar \xi_1(t,\epsilon),\bar \xi_2(t,\epsilon))$ be the solution of (\ref{Eq:rsp}) starting from $\xi^-_0=(\xi^-_{10}, \xi^-_{20})=\psi(x^-_0)$. It is plain that 	$\bar \xi_2(t,\epsilon)=e^{-\frac{1}{\epsilon}t}\xi^-_{20}$. Then consider the following ODE
\begin{align}\label{Eq:rspn}
\left[\begin{matrix}
\dot \xi_1\\
\dot \xi_2
\end{matrix} \right]=  \left[\begin{matrix}
F^*(\xi_1)\\
0
\end{matrix} \right] , 
\end{align}
and let $\xi(t)=(\xi_1(t),\xi_2(t))$ be its solution of (\ref{Eq:rsp}) with the initial point $\xi^+_0=(\xi^+_{10}, \xi^+_{20})=\psi(x^+_0)=\psi\circ\Omega_{E,F}(x^-_0)=\pi\circ\psi(x^-_0)=(\xi^-_{10}, 0)$. Define $\gamma(t,\epsilon)=\bar \xi(t,\epsilon)-\xi(t)$, we have
$$
\dot \gamma(t,\epsilon)=\left[\begin{matrix}
0\\
-\frac{1}{\epsilon}\bar \xi_2(t, \epsilon)
\end{matrix} \right]=\left[\begin{matrix}
0\\
-\frac{1}{\epsilon}e^{-\frac{1}{\epsilon}t}\xi^-_{20}
\end{matrix} \right]
$$
and  $\gamma(0,\epsilon)=\xi^-_0-\xi^+_0=(0,\xi^-_{20})$. It follows that $\gamma(t,\epsilon)=(0,e^{-\frac{1}{\epsilon}t}\xi^-_{20})$.  Moreover, it is not hard to deduce that $\bar x(t,\epsilon)=\psi^{-1}\circ\xi(t,\epsilon)$ and  that $x(t)=\psi^{-1}\circ\xi(t)$. Therefore we have
\begin{align*}
\lim\limits_{\epsilon\to 0} ||\bar x(t,\epsilon)-x(t)||=\lim\limits_{\epsilon\to 0} ||\psi^{-1}\circ \bar \xi(t,\epsilon)-\psi^{-1}\circ \xi(t)||\\{\le}\lim\limits_{\epsilon\to 0} K||\bar \xi(t,\epsilon)- \xi(t)||=\lim\limits_{\epsilon\to 0} K||\gamma(t,\epsilon)||=0.
\end{align*} 
{Note that the inequality ``$\le$''  holds in the above results since $\psi^{-1}$  is a diffeomorphism and thus satisfies the Lipschitz condition for a Lipschitz constant $K$.} 
\end{pf}  
\section{Simulation example}\label{sec:5}
Consider the electrical  circuit shown in Figure \ref{Fig:ec} below,  which consists of a capacitor $C$ and a nonlinear resistor $N$ as the simple circuit  discussed in \cite{SastDeso81,chua1989impasse,RabiRhei02}. A controlled current source  $S$ is additionally connected in parallel with $N$ in order to generate nonlinear terms in $E(x)$ of the DAE model.
\begin{figure}[ht]
	\centering	\begin{circuitikz}[/tikz/circuitikz/bipoles/length=0.7 cm,scale=0.8][european]
		\draw (0,0)-- (0,3)  
		to[capacitor,l=$C$,v_>=$v_{C}{=}z$] (3.5,3);
		\draw  (3.5,0)--(0,0);
		\draw (3.5,3) --(5.5,3) to[american current source, i>=$i_{S}{=}{b(x,y)}{\dot y}$, l_=$S$] (5.5,0)--(3.5,0);  
		\draw (3.5,3) to[variable european
		resistor,l=$N$,v_>=$v_{N}{=}y$,i_>=$i_{N}{=}x$](3.5,0);
	\end{circuitikz}
	\caption{An  electrical circuit with a nonlinear resistor and a controlled current source}
	\label{Fig:ec}
\end{figure}
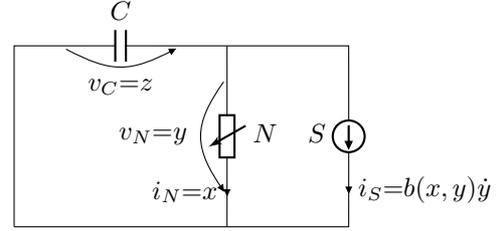
 The relations between the current $i_N=x$ and the voltage $v_N=y$ of the nonlinear resistor $N$ is characterized by the following algebraic equation
$$
0=a(x,y),
$$
and the current $i_S$ of $S$ is equal to $b(x,y)\dot y$, where $a:\mathbb R^2\to \mathbb R$ and $b:\mathbb R^2\to \mathbb R$ are smooth maps  . Using Kirchoff's law, we model the circuit   as a  DAE $\Xi=(E,F)$:  
$$ 
\left[ \begin{smallmatrix}
0&-b(x,y)&C\\
0&0&0\\
0&0&0
\end{smallmatrix}\right]\left[ \begin{smallmatrix}
\dot x\\
\dot y\\
\dot z \end{smallmatrix} \right]   =\left[ \begin{smallmatrix}
x\\
y+z\\
a(x,y)
\end{smallmatrix} \right]. 
$$ 
We consider the following case: $C=1$, $a(x,y)=x-y^2-2y$, $b(x,y)=y$. Let $\eta=(x,y,z)$ and $\eta_p=(0,0,0)$, then the condition  \textbf{(CR)}  is satisfied on $U=\{(x,y,z)\in \mathbb R^3\,|\, y\neq 1\}$. The locally maximal invariant submanifold $M^*$ (around $\eta_p$) is
$
M^*=\left\lbrace \eta\in \mathbb R^3\,|\, y+z=x-y^2-2y=0, y< 1\right\rbrace .
$
Since $\mathcal E=\ker E={\rm span}\{\frac{\partial }{\partial x}, y\frac{\partial }{\partial z}+\frac{\partial }{\partial y}\}$ is involutive and $\Xi$ is of index-1. Then it is possible to find $\psi_1:V\to \mathbb R$, where $V= \left\lbrace \eta\in \mathbb R^3\,|\, y< 1\right\rbrace$, such that ${\rm span}\{\d{\psi_1}\}=\mathcal E^{\bot}$; by solving some first order PDE, we get a solution $\psi_1(\eta)=-\frac{1}{2}y^2+z$. Let $\psi_2(\eta)=y+z$ and $\psi_3=a$, then  the DAE $\Xi$ is locally (on $V$) ex-equivalent to the following DAE represented in the  \textbf{(INWF)}:
\begin{align}\label{Eq:exinwf}
\left[ \begin{smallmatrix}
1&0&0\\
0&0&0\\
0&0&0
\end{smallmatrix}\right]\left[ \begin{smallmatrix}
\dot {\tilde z}\\
\dot {\tilde y}\\
\dot {\tilde x} \end{smallmatrix} \right]   =\left[ \begin{smallmatrix}
-2\tilde z\\
\tilde y\\
\tilde x
\end{smallmatrix} \right]. 
\end{align}
via 
$Q=\left[ \begin{smallmatrix}
1&-2&-1\\
0&1&0\\
0&0&1
\end{smallmatrix}\right]$     and    $\psi(x,y,z)=(\tilde z,\tilde y,\tilde x)=(\psi_1,\psi_2,\psi_3)$. Following (\ref{Eq:ps}) of Definition \ref{Def:sps}, we construct a  singular perturbed system $\Xi_\epsilon$:
\begin{small}
\begin{align*}
Q^{-1}\left[ \begin{smallmatrix}
1&0&0\\
0&-\epsilon&0\\
0&0&-\epsilon
\end{smallmatrix}\right]\frac{\partial \psi}{\partial \eta}\left[ \begin{smallmatrix}
\dot {x}\\
\dot {y}\\
\dot {z} \end{smallmatrix} \right]   =\left[ \begin{smallmatrix}
x\\
y+z\\
x-y^2-2y
\end{smallmatrix} \right]\Rightarrow\Xi_\epsilon: 	 \left[ \begin{smallmatrix}
\dot x\\
\dot y\\
\dot z
\end{smallmatrix}\right] =\left[ \begin{smallmatrix}
f_1(\eta,\epsilon)\\
f_2(\eta,\epsilon)\\
f_3(\eta,\epsilon)
\end{smallmatrix}\right],
\end{align*}
\end{small}
where $f_1(\eta,\epsilon)=-\frac{- x + y (2 + y) - 2 \epsilon (y^2 - 2 z) - 2  (y + z)}{\epsilon}$, $f_2(\eta,\epsilon)=-\frac{y + \epsilon y^2 - 2 \epsilon z + z}{\epsilon + \epsilon y}$, $f_3(\eta,\epsilon)=\frac{\epsilon (y^2 - 2 z) - y (y + z)}{\epsilon (1 + y)}$.
Consider an inconsistent initial point $\eta^-_0=(0,0,0.1)\in V\backslash M^*$, then find the nonlinear consistency projector $\Omega_{E,F}$ to have
$$
\eta^+_0\!=\!\Omega_{E,F}(\eta^-_0)\!=\!\psi^{-1}\circ \pi \circ\psi(\eta^-_0)\!=\!(-0.2,-0.1056,0.1056),
$$
which defines  a  jump $\eta^-_0\to\eta^+_0$ of $\Xi$.
Now we use MATLAB ode45 solver to simulate the solution $\bar \eta(t,\epsilon)=(\bar x(t,\epsilon),\bar y(t,\epsilon),\bar z(t,\epsilon))$ starting from $\eta^-_0$ of the perturbed system $\Xi_\epsilon$ for different values of the perturbation parameter  $\epsilon$ and the $\mathcal C^1$-solution $\eta(t)=(x(t),y(t),z(t))$ of $\Xi$ starting from $\eta^+_0$.
\begin{figure}[ht]
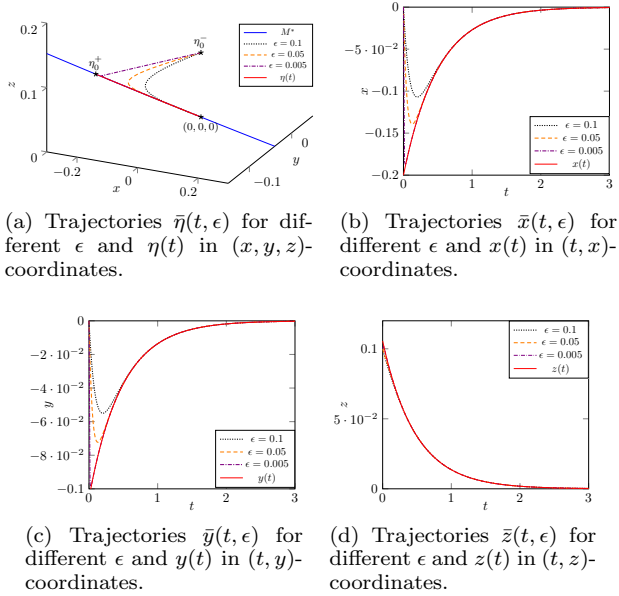

	\centering
	\subfloat[Trajectories $\bar \eta(t,\epsilon) $ for different $\epsilon$ and $\eta(t)$ in $(x,y,z)$-coordinates.]{\input{plot1.tikz}} ~~ 	\subfloat[Trajectories $\bar x(t,\epsilon) $ for different $\epsilon$ and $x(t)$ in $(t,x)$-coordinates.]{\input{plot2.tikz}}\\
		\subfloat[Trajectories $\bar y(t,\epsilon) $ for different $\epsilon$ and $y(t)$ in $(t,y)$-coordinates.]{\input{plot3.tikz}} ~~ 	\subfloat[Trajectories $\bar z(t,\epsilon) $ for different $\epsilon$ and $z(t)$ in $(t,z)$-coordinates.]{\input{plot4.tikz}}
	\caption{ The solutions $\bar \eta(t,\epsilon) $ of   $\Xi_{\epsilon}$ for different   $\epsilon$ and the solution $\eta(t)$ of $\Xi$}
	\label{Fig:ps1} 
\end{figure} 
It can be seen from Figure \ref{Fig:ps1}  that the proposed perturbed system indeed approximates the DAE both for the jump $\eta^-_0\to \eta^+_0$ and for the $\mathcal C^1$-solution $\eta(t)$ starting from $\eta^+_0$ and evolving on $M^*$.  
\section{Conclusions}\label{sec:6} 
In this paper, we  discuss the $\mathcal C^1$-solutions and the jumps from   inconsistent initial points for nonlinear DAEs. First, we propose a normal form called the index-1 nonlinear Weierstrass form \textbf{(INWF)}, which has a simple and decoupled system structure. We show that a nonlinear DAE is   locally externally equivalent to the \textbf{(INWF)} if and only if  the DAE is index-1 and the distribution defined by $\ker E$ is involutive. Then we use the \textbf{(INWF)} to generalize the consistency projector of linear DAEs to the nonlinear case. The generalized nonlinear consistency projector offers a way to solve the consistent initialization problem for nonlinear DAEs. Finally, we propose a system approximation for nonlinear DAEs with jumps via the singular perturbation theory. The results of this paper could be  a nice tool to study hybrid DAE systems involving with switchings since the consistent initialization is a fundamental problem for the solutions of switched nonlinear DAEs.
\begin{small}
\bibliography{bibthesis}   
\end{small}









\end{document}